\documentclass[11pt,twocolumn]{article}
\usepackage[margin=1in]{geometry}
\usepackage{times}
\usepackage{graphicx}
\usepackage{booktabs}
\usepackage{amsmath}
\usepackage{hyperref}
\usepackage{subcaption}
\usepackage{caption}

\title{Democratizing GraphRAG: Linear, CPU-Only Graph Retrieval for Multi-Hop QA}
\author{
Qizhi Wang\\
PingCAP, Data \& AI-Innovation Lab\\
Beijing, China\\
\texttt{qizhi.wang@pingcap.com}
}
\date{}

\begin{document}
\maketitle

\begin{abstract}
GraphRAG systems improve multi-hop retrieval by modeling structure, but many approaches rely on expensive LLM-based graph construction and GPU-heavy inference. We present \textbf{SPRIG} (\textbf{S}eeded \textbf{P}ropagation for \textbf{R}etrieval \textbf{I}n \textbf{G}raphs), a CPU-only, linear-time, token-free GraphRAG pipeline that replaces LLM graph building with lightweight NER-driven co-occurrence graphs and uses Personalized PageRank (PPR) for retrieval. We evaluate on HotpotQA and 2WikiMultiHopQA with BM25, dense retrieval, reciprocal-rank fusion (RRF), and a TF--IDF PPR baseline, focusing on retrieval quality and system efficiency. We provide ablations over seed size, NER choice, and PPR parameters, plus scalability curves and end-to-end QA on a 1{,}000-query subset. Lightweight alias disambiguation and hub pruning reduce query time by 16--28\% with negligible Recall@10 changes. The results characterize when CPU-friendly graph retrieval helps multi-hop recall and when strong lexical hybrids (RRF) are sufficient, outlining a realistic path to democratizing GraphRAG without token costs or GPU requirements.
\end{abstract}

\section{Introduction}
Retrieval-augmented generation (RAG) mitigates hallucinations by grounding LLMs in external evidence, yet dense retrieval can struggle with multi-hop reasoning where evidence spans multiple documents. Graph-based retrieval augments RAG with structured traversal, but many GraphRAG variants rely on LLMs for relation extraction or graph summarization, driving cost and hardware requirements \cite{graphrag,hipporag,hipporag2}. This limits adoption in academic and low-resource settings, where CPU-only constraints and tight memory budgets are common.

This paper studies a pragmatic question: \emph{How far can we go with a linear-time, CPU-only GraphRAG that uses no LLMs during indexing?} We focus on a lightweight entity--document co-occurrence graph and sparse Personalized PageRank (PPR) retrieval. This setup is deliberately simple, but it provides a transparent baseline for ``democratized'' GraphRAG that can be run on commodity hardware, and it surfaces the concrete tradeoffs between retrieval quality and computational cost.

The research significance is twofold. First, it addresses a realistic deployment bottleneck: many academic and industrial teams cannot afford LLM-based graph construction at scale or GPU-based graph inference. Second, by providing a fully reproducible, CPU-only baseline with explicit complexity and memory reporting, we offer a reference point for future work on efficient GraphRAG and green AI.

We target a complementary goal: \textbf{linear-time, CPU-only GraphRAG} that avoids token costs while preserving multi-hop retrieval benefits. Our pipeline builds an entity--document bipartite graph using lightweight NER (SpaCy or regex), then applies Personalized PageRank (PPR) seeded either from query entities or from dense retrieval hits. The design emphasizes linear construction, sparse computation, and practical reproducibility.

Contributions:
\begin{itemize}
  \item \textbf{SPRIG}, a CPU-only, linear-time, token-free GraphRAG pipeline with NER-based co-occurrence graphs.
  \item Lightweight graph refinements for CPU settings: (i) title-alias entity disambiguation (SPRIG-EL), (ii) hub pruning / edge caps (SPRIG-PRUNE), and (iii) explicit entity--document seed mixing (SPRIG-MIX).
  \item GraphHybrid/GraphDense variants that improve Recall@10 over BM25 on both datasets with hub-aware normalization and rank-weighted seeding.
  \item Comparisons against CPU-friendly baselines (RM3, BM25+CE reranker) and an ANN vs.\ exact seeding analysis.
  \item Extensive efficiency curves, latency distributions, and ablations over seed size, seed weighting, hub penalty, and PPR modes.
\end{itemize}

\section{Related Work}
GraphRAG methods use graph structure to improve retrieval and reasoning \cite{graphrag,hipporag,grag,gretriever}. Recent work explores efficient and token-free variants, including LinearRAG and lightweight graph indexing \cite{linearrag,minirag,lightrag,terag}. HippoRAG and TERAG emphasize low-token graph retrieval pipelines; we focus on a retrieval-only, CPU-constrained instantiation and provide empirical baselines under strict memory reporting. These methods reduce indexing cost but often still assume GPU-heavy inference or richer graph construction. We position our work as a practical CPU-only instantiation of the linear GraphRAG philosophy, focusing on retrieval evaluation rather than generation to isolate the indexing and traversal effects.

BM25 remains a strong sparse baseline \cite{bm25}, while dense retrievers (e.g., Sentence-BERT) provide semantic matching \cite{sbert}. Classical pseudo-relevance feedback (RM3) and reranking with cross-encoders remain competitive, CPU-feasible baselines \cite{rm3,bert_rerank}. Multi-hop dense methods such as MDR and beam-style retrieval learn multi-step routing with training-heavy encoders \cite{mdr,beamretrieval}. PPR-based sparse baselines (MixPR) and LLM-assisted graph pipelines (LightPAL) are closely related to our traversal mechanism, though they differ in cost profiles \cite{mixpr,lightpal}. Classical text-graph ranking (TextRank, LexRank) and topic-sensitive PageRank on term/document graphs provide early random-walk retrieval baselines that our approach echoes at the entity--document level \cite{textrank,lexrank,topicsensitivepagerank}. Agentic and graph-expansion systems (GEAR/SyncGE, HopRAG) and efficient iterative retrievers also target multi-hop evidence discovery but rely on heavier models or token budgets \cite{gear,hoprag,effrag}. GNN-based KG retrievers such as GFM-RAG train graph encoders for multi-hop retrieval \cite{gfmrag}. Recent CPU-friendly dense retrievers such as LADR (lexically accelerated dense retrieval with proximity graphs) \cite{ladr} and MoPo (multi-hop dense retrieval via posterior distillation) \cite{mopo} are strong non-LLM alternatives; we summarize their key requirements qualitatively in Appendix Table~\ref{tab:qualitative-compare}. We compare to their principles conceptually and focus on CPU-only feasibility under a strict 4~GB memory budget.

\section{Method}
\subsection{Pipeline Overview}
Our pipeline follows a simple, deterministic sequence: (1) normalize text; (2) extract entities with lightweight NER; (3) build an entity--document bipartite graph with TF--IDF weighting; (4) create a seed distribution from the query (entities or top-$k$ BM25/dense passages) with optional rank-based weighting; (5) run PPR by power iteration or a push-based approximation; (6) rank passages by PPR scores. This design keeps indexing linear in corpus size and inference bounded by sparse matrix-vector multiplications.

\subsection{Graph Construction (Linear-Time)}
We construct an entity--document bipartite graph. For each document passage $d$, we extract entities $e$ via lightweight NER (SpaCy or a regex heuristic). The regex heuristic extracts 1--4 token capitalized spans using the pattern \texttt{\textbackslash b[A-Z][a-z]+(\textbackslash s+[A-Z][a-z]+)\{0,3\}\textbackslash b}. We optionally normalize entities with a \texttt{simple} rule (lowercasing and stripping non-alphanumeric characters) to reduce surface-form variation. We add weighted edges
\[
w_{e,d} = \mathrm{tf}(e,d)\cdot \log\left(\frac{N+1}{\mathrm{df}(e)+1}\right) + 1,
\]
where $N$ is the number of passages and $\mathrm{df}(e)$ is entity document frequency. The adjacency matrix $W$ is row-normalized to obtain the transition matrix $P = D^{-1}W$ (row-stochastic). The process is linear in the number of entity mentions and does not invoke LLMs. We also build a TF--IDF term--document graph (MixPR-like baseline) by replacing entities with token terms, filtering terms with $\mathrm{df}<3$ or $\mathrm{df}>0.1N$.

We do \emph{not} perform full entity linking or coreference resolution; however, we add a lightweight title-alias disambiguation (SPRIG-EL) by mapping normalized mentions to unambiguous document-title aliases (e.g., stripping parenthetical disambiguators). This yields a zero-KB, CPU-only alias map that reduces surface-form variation without external supervision. We partially mitigate remaining ambiguity by IDF weighting, optional entity normalization, and hub downweighting: we scale document$\rightarrow$entity edges by $\mathrm{df}(e)^{-p}$ (with $p\ge 0$), which reduces the influence of high-degree entity hubs. For efficiency we optionally prune hubs by removing the top-$x\%$ entities by $\mathrm{df}$ and/or capping per-entity outdegree to the top-$L$ TF--IDF edges (SPRIG-PRUNE), reducing $E$ and PPR cost.

\subsection{Retrieval via PPR}
Given a query, we identify query entities (for \textit{Graph}) or use BM25/dense retrieval to select seed passages (for \textit{GraphHybrid/GraphDense}). We run PPR with damping factor $\alpha$ for $T$ iterations (power iteration) or use a push-based approximation with residual threshold $\epsilon$:
\[
r = \alpha s + (1-\alpha) P^\top r,
\]
where $s$ is the seed distribution over entity/passages. Here $P$ is the row-stochastic transition matrix over the combined entity+document nodes; we use the standard PPR update with $P^\top$ and do not use a two-step ($P^2$) variant. On a bipartite graph, the random walk naturally alternates between entities and documents, and we rank passages by the document slice of $r$. For Graph we seed only entity nodes (uniform over matched query entities, optionally downweighted by $\mathrm{df}(e)^{-q}$). For GraphHybrid/GraphDense we construct separate seed vectors $s_e$ (entities) and $s_d$ (seed passages) and mix them as
\[
s = \mathrm{norm}\left(\alpha_{\text{mix}} s_e + (1-\alpha_{\text{mix}}) s_d\right),
\]
with $\mathrm{norm}$ denoting $L_1$ normalization over nonzero entries. The default is ``mass-proportional'' mixing ($\alpha_{\text{mix}}$ implicit from raw $L_1$ mass), while an adaptive option uses $\alpha_{\text{mix}}=\frac{n_e+1}{n_e+n_d+2}$ with $n_e,n_d$ the number of nonzero seeds (Laplace-smoothed). If either partition is empty, we fall back to the other. We add seed mass on the top-$k$ BM25/dense passages using raw, softmax, or rank-based weighting; we use rank weighting in main results. If a query yields no recognized entities, we fall back to a uniform distribution over passages; a BM25 top-1 fallback is evaluated in Appendix Table~\ref{tab:graph-fallback}.

\subsection{Complexity and Memory}
Graph construction is $O(M)$ where $M$ is the number of entity (or term) mentions; edges $E$ scale with mentions. Each PPR query costs $O(T\cdot E)$ using sparse matrix-vector multiplication, and memory is $O(E+V)$ for sparse adjacency and node scores. Push-based PPR replaces fixed iterations with residual-driven updates that can reduce work on sparse queries. Hub pruning or edge caps reduce $E$ directly, lowering query time while preserving most recall in practice. Appendix Table \ref{tab:graph-stats} shows that entity graphs have heavy-tailed degree distributions (hub entities), motivating degree-aware normalization or filtering. We report peak RSS and index/query RSS in experiments to verify the 4~GB constraint.

\section{Experimental Setup}
\subsection{Datasets}
We evaluate on HotpotQA and 2WikiMultiHopQA \cite{hotpotqa,twowiki}. For HotpotQA validation, we use 7{,}405 queries and 66{,}581 passages; for 2WikiMultiHopQA validation, we use 10{,}000 queries and 45{,}902 passages. We use dataset-provided passages as retrieval units without additional chunking to keep preprocessing deterministic and to reflect the standard multi-hop QA setting.

\subsection{Evaluation Metrics}
We report Recall@K, Hit@K, and MRR for retrieval. Recall@K measures the fraction of gold supporting passages retrieved in top-$K$, while Hit@K is the fraction of questions with at least one gold passage in top-$K$. MRR is computed as $ \frac{1}{|Q|}\sum_{q \in Q} \frac{1}{\min\{r : d_r \in G_q\}}$, with 0 if no relevant passage is retrieved; ties are resolved by the system's ranked order. Because MRR uses only the rank of the first relevant passage, it can be high even when Recall@K is moderate under multi-label ground truth. We additionally measure end-to-end QA using a small extractive model (DistilBERT QA) on 1{,}000-query subsets to test downstream impact.

\subsection{Baselines}
We compare: (1) BM25, (2) BM25+RM3 pseudo-relevance feedback \cite{rm3}, (3) BM25-2step entity expansion (two-stage lexical routing with entity-based query expansion), (4) Dense (BAAI/bge-small-en-v1.5), (5) RRF (BM25+dense fusion), (6) BM25+CE (TinyBERT cross-encoder reranking top-100) \cite{bert_rerank,tinybert}, (7) TF--IDF Graph PPR (MixPR-like), (8) Graph (query-entity seeded), (9) GraphHybrid (BM25 seeds), and (10) GraphDense (dense seeds). RRF uses $1/(k+r)$ scoring with $k{=}60$, a standard robust fusion setting. Appendix Table \ref{tab:graph-rrf} adds RRF variants (GraphRRF, RRF+PPR score fusion, and RRF+CE) to disentangle seed enrichment from score-level ensembling.

\subsection{Graph and PPR Settings}
Unless otherwise noted, we set min\_entity\_len=2, min\_entity\_df=1, max\_entity\_df\_ratio=1.0, and row normalization for the graph transition. The TF--IDF term graph uses min\_df=3 and max\_df\_ratio=0.1 (tuned on validation subsets). PPR uses $\alpha{=}0.15$ and max\_iter=5; we use push-based PPR for HotpotQA and power iteration for 2Wiki (Appendix). Seed sizes are $k{=}5$ (GraphDense) and $k{=}10$ (GraphHybrid) for HotpotQA, and $k{=}3$ (GraphDense) and $k{=}5$ (GraphHybrid) for 2Wiki. We use rank-based seed weighting and hub downweighting with $p{=}0.5$, and downweight query-entity seeds by $\mathrm{df}(e)^{-q}$ with $q{=}0.5$ (HotpotQA) or $q{=}1.0$ (2Wiki). Entity normalization uses \texttt{simple} (HotpotQA) and \texttt{lower} (2Wiki).

\paragraph{QTime definition.} QTime reports end-to-end per-query time for each method. For graph variants that require external seeding (GraphHybrid, GraphDense, GraphRRF, and RRF+PPR fusion), QTime includes the seeding stage (BM25/dense/RRF) plus PPR traversal; we additionally log seed vs.\ traversal time separately (Appendix Table~\ref{tab:graph-timing-breakdown}).

\subsection{Hyperparameter Tuning Protocol}
We tune graph/PPR hyperparameters on a 500-query subset of each validation set (seed size $k$, seed weighting, hub penalty $p$, entity normalization, seed-entity downweighting $q$, and PPR mode). The reported main-table results are computed on the full validation splits with the selected configuration. This may introduce mild validation overfitting; we mitigate this by reporting extensive ablations, paired bootstrap confidence intervals, and a robustness check that removes the tuning subset from evaluation (Appendix Table~\ref{tab:robustness-no-tune}).

\subsection{Environment}
All experiments run on CPU-only hardware with a 4 GB RAM budget, and we report peak RSS plus index/query RSS. SpaCy NER uses \texttt{en\_core\_web\_sm} (v3.8.0; spaCy 3.8.11); on our CPU, graph construction costs roughly 9--10\,ms/doc with SpaCy versus $\sim$0.02\,ms/doc with the regex heuristic. Dense retrieval uses BAAI/bge-small-en-v1.5 in the main tables and HNSW for approximate nearest neighbors (ANN) with $M{=}32$, \texttt{efConstruction}{=}200, and \texttt{efSearch}{=}64 unless noted; we report index time, total query time, and per-query latency distributions for each method. Appendix Table~\ref{tab:dense-model-sensitivity} compares additional dense models (including all-MiniLM-L6-v2 and e5-small). We also quantify ANN versus exact dense seeding for GraphDense on a 2k-query subset (Appendix Table~\ref{tab:graph-dense-seeding}) and report HNSW tuning (Appendix Table~\ref{tab:graph-dense-ann-tuning}). The reranking baseline uses a TinyBERT cross-encoder on the top-100 BM25 candidates (CPU-only).

\section{Main Results}
Tables \ref{tab:hotpot-results} and \ref{tab:twowiki-results} summarize retrieval quality and efficiency (Dense uses bge-small; Appendix Table~\ref{tab:dense-model-sensitivity} provides model sensitivity). RRF remains a strong hybrid baseline, while the CPU-friendly reranker (BM25+CE) delivers strong MRR at higher latency. GraphHybrid consistently improves over BM25 and is competitive with RRF: it surpasses RRF on 2Wiki but falls short on HotpotQA, illustrating the dependence on seed quality and graph noise. GraphDense is competitive and slightly higher than GraphHybrid in R@10 on both datasets, while query-entity-only Graph remains weaker. RRF-seeded PPR (GraphRRF; Appendix Table \ref{tab:graph-rrf}) further improves Recall@10 over RRF, showing that stronger seeds plus traversal can yield gains.

We report paired bootstrap confidence intervals for $\Delta$R@10 against BM25 and RRF for all methods (Appendix Tables \ref{tab:sig-hotpot-bm25}--\ref{tab:sig-2wiki-rrf}). These paired tests are computed from the same predictions used in the main tables (merged across method-specific runs) to avoid label or run mismatches. Default HNSW settings can yield noticeable gaps versus exact seeding (Appendix Table \ref{tab:graph-dense-seeding}); tuning $M$/\texttt{efSearch} narrows the gap on 2k-query subsets (Appendix Table \ref{tab:graph-dense-ann-tuning}), and full-validation results are comparable to default settings (Table \ref{tab:graph-dense-tuned-full}), indicating that ANN recall is a meaningful driver of gains.

\begin{table}[t]
\centering
\caption{HotpotQA validation results (7{,}405 queries). QTime is total query time in seconds.}
\label{tab:hotpot-results}
\resizebox{\linewidth}{!}{\begin{tabular}{lrrrrr}
Method & R@5 & R@10 & Hit@10 & MRR & QTime \\
\hline
BM25 & 0.631 & 0.742 & 0.960 & 0.784 & 444.4 \\
BM25+RM3 & 0.531 & 0.675 & 0.929 & 0.662 & 1292.8 \\
BM25-2step & 0.617 & 0.729 & 0.944 & 0.732 & 965.4 \\
Dense & 0.758 & 0.811 & 0.958 & 0.878 & 71.2 \\
RRF & 0.764 & 0.851 & 0.988 & 0.865 & 518.1 \\
BM25+CE & 0.731 & 0.810 & 0.983 & 0.881 & 1358.1 \\
TF-IDF Graph & 0.290 & 0.419 & 0.693 & 0.319 & 8883.1 \\
Graph & 0.374 & 0.464 & 0.656 & 0.448 & 467.9 \\
GraphHybrid & 0.655 & 0.775 & 0.964 & 0.778 & 954.4 \\
GraphDense & 0.782 & 0.844 & 0.968 & 0.875 & 582.7 \\
\end{tabular}
}
\end{table}

\begin{table}[t]
\centering
\caption{2WikiMultiHopQA validation results (10{,}000 queries).}
\label{tab:twowiki-results}
\resizebox{\linewidth}{!}{\begin{tabular}{lrrrrr}
Method & R@5 & R@10 & Hit@10 & MRR & QTime \\
\hline
BM25 & 0.579 & 0.643 & 0.972 & 0.819 & 303.1 \\
BM25+RM3 & 0.486 & 0.585 & 0.927 & 0.674 & 900.8 \\
BM25-2step & 0.569 & 0.653 & 0.949 & 0.778 & 618.3 \\
Dense & 0.579 & 0.609 & 0.930 & 0.885 & 86.0 \\
RRF & 0.654 & 0.697 & 0.993 & 0.914 & 405.0 \\
BM25+CE & 0.627 & 0.676 & 0.985 & 0.891 & 2100.2 \\
TF-IDF Graph & 0.275 & 0.367 & 0.648 & 0.331 & 380.9 \\
Graph & 0.318 & 0.357 & 0.516 & 0.430 & 323.8 \\
GraphHybrid & 0.622 & 0.743 & 0.964 & 0.839 & 709.0 \\
GraphDense & 0.681 & 0.747 & 0.952 & 0.901 & 490.9 \\
\end{tabular}
}
\end{table}

\begin{table}[t]
\centering
\caption{Effect of tuned HNSW on full-validation GraphDense.}
\label{tab:graph-dense-tuned-full}
\resizebox{\linewidth}{!}{\begin{tabular}{llrrrr}
Dataset & Seeding & R@10 & Hit@10 & MRR & QTime \\
\hline
HotpotQA & ANN (default) & 0.844 & 0.968 & 0.875 & 582.7 \\
HotpotQA & ANN (tuned) & 0.844 & 0.968 & 0.875 & 582.7 \\
2WikiMultiHopQA & ANN (default) & 0.747 & 0.952 & 0.901 & 490.9 \\
2WikiMultiHopQA & ANN (tuned) & 0.747 & 0.952 & 0.901 & 490.9 \\
\end{tabular}
}
\end{table}

End-to-end QA on 1{,}000-query subsets (DistilBERT QA) shows modest gains; we provide detailed EM/F1 in Appendix Table \ref{tab:qa-results}. This indicates that retrieval improvements do not always translate into large downstream QA gains under a small extractive model.

\section{Ablations}
We ablate seed size ($k$), seed weighting, NER choice (SpaCy vs.\ regex), hub penalty $p$, seed-entity downweighting $q$, and PPR parameters (mode, $\alpha$, max\_iter) on 500-query subsets for both datasets due to CPU and memory constraints. Across datasets, rank-based seeding outperforms raw or softmax weighting, and modest hub downweighting ($p \approx 0.5$) improves robustness. Regex NER is orders of magnitude faster for indexing and can be competitive for Graph-only retrieval, but we retain SpaCy in main tables for a standard NER baseline. Push-based PPR slightly helps HotpotQA while power iteration is competitive on 2Wiki. We additionally evaluate RRF variants (GraphRRF, RRF+PPR fusion, and RRF+CE) in Table \ref{tab:graph-rrf} and a BM25 fallback for entity-missing queries in Appendix Table \ref{tab:graph-fallback}. Detailed plots and top configurations are provided in the Appendix.

\paragraph{Entity extraction sensitivity.} We analyze a lightweight proxy for NER quality by measuring (i) query entity coverage and (ii) whether extracted entities match any gold titles under the same normalization/aliasing. Appendix Table~\ref{tab:ner-proxy} shows that Graph-only retrieval is most sensitive to missed entities, while GraphHybrid is more robust because BM25 seeding compensates for NER errors.

\subsection{SPRIG-EL/PRUNE/MIX Enhancements}
Table \ref{tab:graph-enhancements-full} compares GraphHybrid with and without lightweight enhancements on the full validation sets. We set hub pruning to remove the top 1\% entities by $\mathrm{df}$ and use adaptive mixing ($\alpha_{\text{mix}}=\frac{n_e+1}{n_e+n_d+2}$); title-alias disambiguation is built from corpus titles (no external KB). Hub pruning and explicit seed mixing reduce query time substantially (HotpotQA: 485.3\,s $\rightarrow$ 350.5\,s; 2Wiki: 367.9\,s $\rightarrow$ 308.5\,s) with negligible Recall@10 changes. Appendix Table \ref{tab:graph-enhancements-q1000} reports q1000 ablations showing that hub pruning drives most of the latency gains, while title-alias linking yields small but consistent effects. These refinements provide a practical CPU-time lever without introducing heavy models.

We further quantify hub pruning coverage by measuring how often pruned entities overlap with gold titles. Appendix Table~\ref{tab:hub-pruning} shows that removing the top 1\% entity hubs eliminates only a small fraction of gold-title mentions, suggesting pruning mostly removes generic hubs rather than rare bridge entities.

\begin{table}[t]
\centering
\caption{Effect of lightweight enhancements on full validation (GraphHybrid).}
\label{tab:graph-enhancements-full}
\resizebox{\linewidth}{!}{\begin{tabular}{lrrrr}
Variant & Hotpot R@10 & Hotpot QTime (s) & 2Wiki R@10 & 2Wiki QTime (s) \\
\hline
Base & 0.775 & 485.3 & 0.743 & 367.9 \\
+ALL & 0.773 & 350.5 & 0.743 & 308.5 \\
\end{tabular}
}
\end{table}

\begin{table}[t]
\centering
\caption{RRF variants and graph hybrids: GraphRRF, RRF+PPR fusion, and RRF+CE.}
\label{tab:graph-rrf}
\resizebox{\linewidth}{!}{\begin{tabular}{llrrrr}
Dataset & Method & R@10 & Hit@10 & MRR & QTime \\
\hline
HotpotQA & GraphDense & 0.844 & 0.968 & 0.875 & 582.7 \\
HotpotQA & GraphHybrid & 0.775 & 0.964 & 0.778 & 954.4 \\
HotpotQA & RRF & 0.851 & 0.988 & 0.865 & 518.1 \\
HotpotQA & GraphRRF & 0.867 & 0.989 & 0.852 & 1048.8 \\
HotpotQA & RRF+PPR (fusion) & 0.782 & 0.961 & 0.770 & 994.5 \\
HotpotQA & RRF+CE & 0.846 & 0.989 & 0.887 & 638.4 \\
2WikiMultiHopQA & GraphDense & 0.747 & 0.952 & 0.901 & 490.9 \\
2WikiMultiHopQA & GraphHybrid & 0.743 & 0.964 & 0.839 & 709.0 \\
2WikiMultiHopQA & RRF & 0.697 & 0.993 & 0.914 & 405.0 \\
2WikiMultiHopQA & GraphRRF & 0.794 & 0.990 & 0.912 & 814.6 \\
2WikiMultiHopQA & RRF+PPR (fusion) & 0.602 & 0.952 & 0.867 & 761.0 \\
2WikiMultiHopQA & RRF+CE & 0.701 & 0.993 & 0.901 & 701.3 \\
\end{tabular}
}
\end{table}

\section{Efficiency and Scalability}
\begin{figure}[t]
\centering
\begin{subfigure}[b]{0.48\linewidth}
\includegraphics[width=\linewidth]{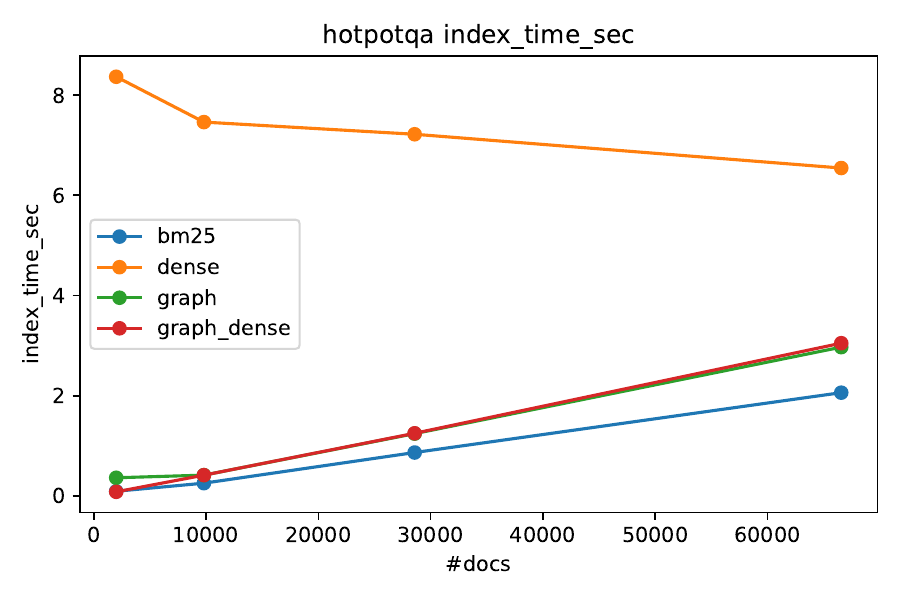}
\caption{HotpotQA index}
\end{subfigure}
\hfill
\begin{subfigure}[b]{0.48\linewidth}
\includegraphics[width=\linewidth]{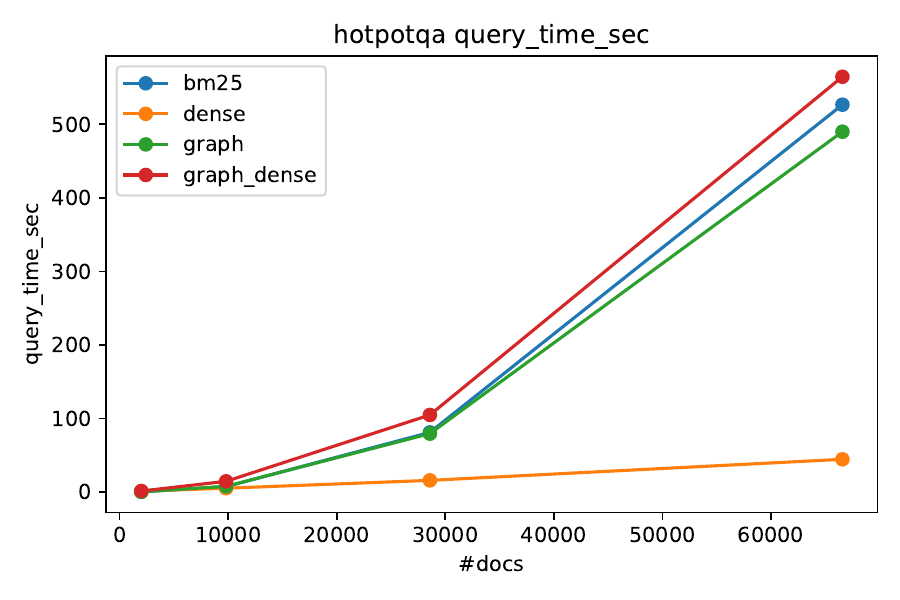}
\caption{HotpotQA query}
\end{subfigure}

\begin{subfigure}[b]{0.48\linewidth}
\includegraphics[width=\linewidth]{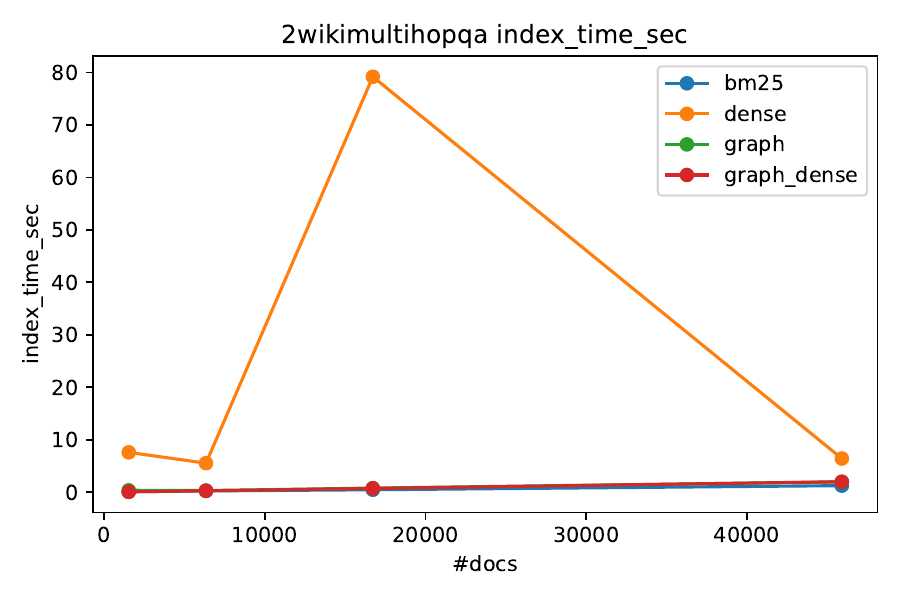}
\caption{2Wiki index}
\end{subfigure}
\hfill
\begin{subfigure}[b]{0.48\linewidth}
\includegraphics[width=\linewidth]{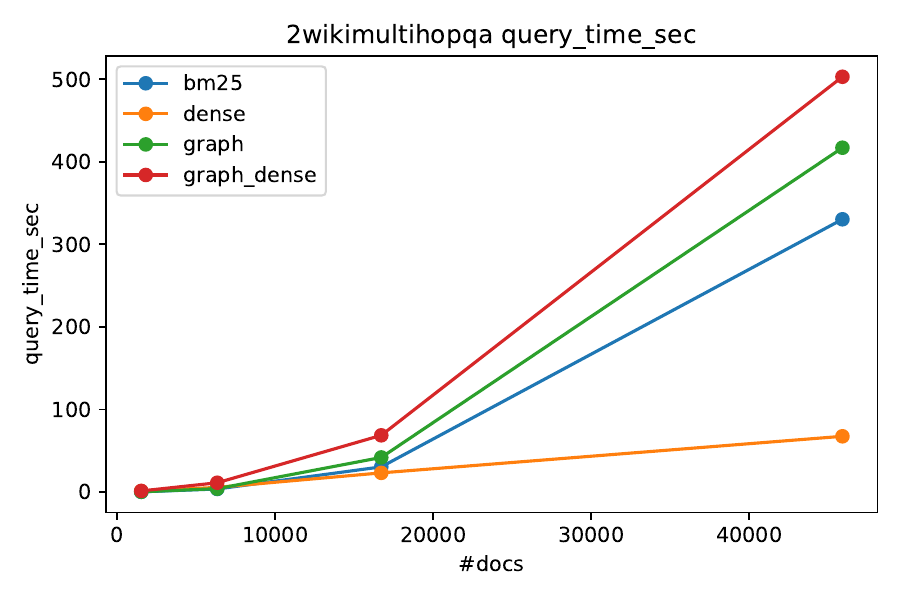}
\caption{2Wiki query}
\end{subfigure}
\caption{Efficiency curves for index and query time.}
\label{fig:efficiency}
\end{figure}

We measure index time and query time across dataset sizes, shown in Figure \ref{fig:efficiency}. Indexing scales roughly linearly with corpus size, consistent with linear-time graph construction; small non-monotonic fluctuations (notably dense index time) stem from HNSW randomness, caching/IO effects, and OS scheduling since each size is run once. To normalize for corpus size, we report per-document index time ranges in Appendix Table \ref{tab:efficiency-per-doc}. Query time grows with graph density and the number of PPR iterations; graph variants that rely on external seeding (GraphHybrid/GraphRRF) are the most expensive among graph methods, while the cross-encoder reranker is the most expensive overall. We additionally report per-query latency distributions (Appendix) and peak/index/query memory usage (RSS) under 4~GB (Appendix Table \ref{tab:rss-stats}).

\section{Discussion}
Graph-based retrieval recovers multi-hop evidence beyond lexical matching, but performance is sensitive to seed quality and graph noise. RRF provides strong lexical+semantic fusion, while the CPU-friendly reranker (BM25+CE) offers strong MRR at a substantial latency cost. GraphHybrid improves over BM25 and is competitive with RRF; GraphDense is slightly stronger on R@10 in both datasets, and query-entity-only Graph remains weaker. RRF-seeded PPR (GraphRRF) further improves Recall@10 over RRF (Table \ref{tab:graph-rrf}), suggesting that richer seeds plus traversal can yield additional gains. ANN tuning affects GraphDense: the 2k-query sweeps show sensitivity to $M$ and \texttt{efSearch} (Appendix Table \ref{tab:graph-dense-ann-tuning}), while full-validation results remain comparable to default settings (Table \ref{tab:graph-dense-tuned-full}). Ablations indicate that rank-based seeding and modest hub downweighting improve robustness; push-based PPR helps on HotpotQA while power iteration is competitive on 2Wiki. Query-time costs remain higher than dense-only retrieval, indicating a tradeoff: graph traversal can improve multi-hop recall yet requires further optimization for latency-sensitive applications. Our lightweight pruning and explicit seed mixing reduce query time by roughly 16--28\% on full validation with negligible Recall@10 changes (Table \ref{tab:graph-enhancements-full}), providing a practical CPU-time lever without heavier models. RM3 underperforms BM25 on these datasets, suggesting that classic PRF is not consistently reliable for multi-hop retrieval. A simple BM25 fallback improves Graph-only queries without entities (Appendix Table \ref{tab:graph-fallback}), but more principled entity linking remains important.

The TF--IDF graph baseline performs poorly relative to entity graphs, suggesting that explicit entity co-occurrence provides a more effective topology for multi-hop navigation than term-level edges. This supports the use of lightweight NER even without full entity linking.

\section{Limitations}
We evaluate retrieval on two multi-hop QA datasets; end-to-end QA is evaluated on 1{,}000-query subsets with a small extractive model. We do not perform full entity linking or coreference resolution; our title-alias disambiguation is intentionally minimal and only resolves unambiguous aliases, so high-degree entities and residual ambiguity can still introduce noise (Appendix reports degree statistics). We quantify NER sensitivity with a lightweight proxy based on query entity coverage and gold-title matches (Appendix Table~\ref{tab:ner-proxy}), but we do not provide full human-labeled NER precision/recall. Approximate dense indexing (HNSW) trades recall for speed; default settings can underperform exact seeding, and tuning reduces that gap on subsets (Appendix Table~\ref{tab:graph-dense-seeding}), while full-validation results remain similar (Table~\ref{tab:graph-dense-tuned-full}). We test one dense encoder in main results and include small-model sensitivity in the Appendix (Table~\ref{tab:dense-model-sensitivity}); broader model choices or stronger cross-encoders may shift the accuracy--latency frontier.
Our hyperparameters are tuned on validation subsets rather than a held-out tuning split, which may introduce mild overfitting; we report a robustness check that removes the tuning subset from evaluation (Appendix Table~\ref{tab:robustness-no-tune}). We do not re-implement training-heavy baselines (e.g., MDR) or recent non-LLM multi-hop dense retrievers such as LADR and MoPo \cite{ladr,mopo} under identical CPU-only constraints; instead, we provide a qualitative positioning in Appendix Table~\ref{tab:qualitative-compare}. Similarly, we do not provide head-to-head runs against token-efficient GraphRAG pipelines (e.g., LinearRAG, HippoRAG/TERAG) beyond conceptual positioning. Extending to additional datasets (e.g., MuSiQue) and stronger readers are left for future work.

\section{Conclusion}
We present a CPU-only, linear-time GraphRAG pipeline that avoids LLM-based graph construction while improving multi-hop retrieval quality. Our results and ablations suggest that simple co-occurrence graphs plus PPR provide a practical and reproducible path toward democratizing GraphRAG. We will release code and scripts to reproduce the 4~GB CPU-only setting, including indexing and PPR parameters.
We additionally show that lightweight title-alias disambiguation and hub pruning reduce query-time costs with negligible Recall@10 changes, providing an actionable efficiency knob for CPU deployments.

\bibliographystyle{plain}
\bibliography{refs}

\appendix
\renewcommand{\thetable}{A.\arabic{table}}
\renewcommand{\thefigure}{A.\arabic{figure}}
\setcounter{table}{0}
\setcounter{figure}{0}
\section{Ablation Details}
Tables \ref{tab:ablation-hotpot} and \ref{tab:ablation-2wiki} list the top configurations from the 500-query ablation subsets. Read them as ranked configurations by Recall@10; columns report the NER type, seed size $k$, $(\alpha,\text{iter})$, and the resulting retrieval metrics and query time.

\begin{table}[htbp]
\centering
\caption{HotpotQA ablation (q500 subset): top configurations by Recall@10.}
\label{tab:ablation-hotpot}
\resizebox{\linewidth}{!}{\begin{tabular}{lrrrrrrrr}
Dataset & NER & k & $\alpha$ & it & R@5 & R@10 & MRR & QTime \\
\hline
hotpotqa & spacy & 5 & 0.15 & 5 & 0.647 & 0.770 & 0.674 & 8.4 \\
hotpotqa & spacy & 5 & 0.10 & 5 & 0.639 & 0.769 & 0.661 & 8.2 \\
hotpotqa & spacy & 5 & 0.20 & 5 & 0.652 & 0.768 & 0.684 & 8.6 \\
hotpotqa & spacy & 5 & 0.15 & 20 & 0.644 & 0.764 & 0.647 & 11.8 \\
hotpotqa & spacy & 5 & 0.20 & 20 & 0.647 & 0.764 & 0.657 & 12.0 \\
hotpotqa & spacy & 5 & 0.20 & 10 & 0.645 & 0.763 & 0.651 & 9.6 \\
hotpotqa & spacy & 3 & 0.20 & 5 & 0.668 & 0.762 & 0.725 & 8.4 \\
hotpotqa & spacy & 3 & 0.10 & 5 & 0.666 & 0.761 & 0.702 & 8.4 \\
hotpotqa & spacy & 3 & 0.15 & 5 & 0.671 & 0.760 & 0.716 & 8.3 \\
hotpotqa & spacy & 3 & 0.20 & 20 & 0.666 & 0.760 & 0.709 & 12.1 \\
\end{tabular}
}
\end{table}

\begin{table}[htbp]
\centering
\caption{2WikiMultiHopQA ablation (q500 subset): top configurations by Recall@10.}
\label{tab:ablation-2wiki}
\resizebox{\linewidth}{!}{\begin{tabular}{lrrrrrrrr}
Dataset & NER & k & $\alpha$ & it & R@5 & R@10 & MRR & QTime \\
\hline
2wikimultihopqa & spacy & 3 & 0.15 & 5 & 0.605 & 0.690 & 0.762 & 15.0 \\
2wikimultihopqa & spacy & 3 & 0.10 & 5 & 0.607 & 0.690 & 0.753 & 4.9 \\
2wikimultihopqa & spacy & 3 & 0.15 & 20 & 0.598 & 0.689 & 0.724 & 6.9 \\
2wikimultihopqa & spacy & 3 & 0.20 & 5 & 0.604 & 0.689 & 0.767 & 5.2 \\
2wikimultihopqa & spacy & 3 & 0.10 & 20 & 0.594 & 0.686 & 0.714 & 6.7 \\
2wikimultihopqa & spacy & 3 & 0.20 & 20 & 0.599 & 0.686 & 0.733 & 7.1 \\
2wikimultihopqa & spacy & 3 & 0.20 & 10 & 0.597 & 0.684 & 0.725 & 6.0 \\
2wikimultihopqa & spacy & 3 & 0.15 & 10 & 0.596 & 0.684 & 0.709 & 6.0 \\
2wikimultihopqa & spacy & 3 & 0.10 & 10 & 0.595 & 0.680 & 0.694 & 5.8 \\
2wikimultihopqa & spacy & 5 & 0.10 & 5 & 0.539 & 0.664 & 0.718 & 5.1 \\
\end{tabular}
}
\end{table}

\begin{table}[htbp]
\centering
\caption{SPRIG-EL/PRUNE/MIX ablation on q1000 subsets (GraphHybrid).}
\label{tab:graph-enhancements-q1000}
\resizebox{\linewidth}{!}{\begin{tabular}{lrrrr}
Variant & Hotpot R@10 & Hotpot QTime (s) & 2Wiki R@10 & 2Wiki QTime (s) \\
\hline
Base & 0.855 & 33.2 & 0.838 & 8.6 \\
+EL & 0.855 & 31.3 & 0.841 & 8.9 \\
+PRUNE & 0.854 & 22.5 & 0.838 & 6.0 \\
+MIX & 0.850 & 32.9 & 0.835 & 8.3 \\
+ALL & 0.848 & 23.7 & 0.840 & 7.9 \\
\end{tabular}
}
\end{table}

\section{Graph Statistics and Latency}
Tables \ref{tab:graph-stats}--\ref{tab:latency-stats} provide structural and runtime context, and Table \ref{tab:graph-timing-breakdown} separates seeding from PPR traversal. \textbf{Reading tip:} focus on the entity-degree p95 values to assess hub severity and compare p50/p95 latency to understand tail behavior.
\begin{table}[htbp]
\centering
\caption{Graph statistics (nodes, edges, and p95 degrees).}
\label{tab:graph-stats}
\resizebox{\linewidth}{!}{\begin{tabular}{llrrrr}
Dataset & Graph & Nodes & Edges & Deg$_{p95}$ (E) & Deg$_{p95}$ (D) \\
\hline
HotpotQA & entity\_graph & 335477 & 592345 & 5 & 18 \\
HotpotQA & term\_graph & 102465 & 3030059 & 298 & 88 \\
2WikiMultiHopQA & entity\_graph & 225218 & 376632 & 4 & 23 \\
2WikiMultiHopQA & term\_graph & 72329 & 1683556 & 223 & 113 \\
\end{tabular}
}
\end{table}

\begin{table}[htbp]
\centering
\caption{Per-query latency distribution (seconds).}
\label{tab:latency-stats}
\resizebox{\linewidth}{!}{\begin{tabular}{llrrr}
Dataset & Method & p50 (s) & p95 (s) & p99 (s) \\
\hline
HotpotQA & BM25-2step & 0.126 & 0.200 & 0.247 \\
HotpotQA & BM25+RM3 & 0.169 & 0.237 & 0.294 \\
HotpotQA & BM25 & 0.057 & 0.104 & 0.136 \\
HotpotQA & Dense & 0.009 & 0.011 & 0.017 \\
2WikiMultiHopQA & BM25-2step & 0.062 & 0.088 & 0.098 \\
2WikiMultiHopQA & BM25+RM3 & 0.089 & 0.114 & 0.124 \\
2WikiMultiHopQA & BM25 & 0.030 & 0.048 & 0.056 \\
2WikiMultiHopQA & Dense & 0.008 & 0.010 & 0.013 \\
HotpotQA & GraphDense & 0.077 & 0.099 & 0.114 \\
HotpotQA & RRF+PPR (fusion) & 0.129 & 0.198 & 0.237 \\
HotpotQA & GraphRRF & 0.138 & 0.194 & 0.228 \\
HotpotQA & GraphHybrid & 0.126 & 0.180 & 0.212 \\
HotpotQA & TF-IDF Graph & 1.165 & 1.893 & 2.301 \\
HotpotQA & Graph & 0.061 & 0.104 & 0.126 \\
2WikiMultiHopQA & GraphRRF & 0.080 & 0.104 & 0.124 \\
2WikiMultiHopQA & RRF+PPR (fusion) & 0.075 & 0.099 & 0.114 \\
2WikiMultiHopQA & GraphDense & 0.048 & 0.056 & 0.076 \\
2WikiMultiHopQA & Graph & 0.036 & 0.042 & 0.050 \\
2WikiMultiHopQA & TF-IDF Graph & 0.037 & 0.043 & 0.052 \\
2WikiMultiHopQA & GraphHybrid & 0.070 & 0.092 & 0.107 \\
HotpotQA & RRF & 0.067 & 0.116 & 0.147 \\
HotpotQA & RRF+CE & 0.083 & 0.135 & 0.168 \\
HotpotQA & BM25+CE & 0.178 & 0.255 & 0.309 \\
2WikiMultiHopQA & RRF & 0.040 & 0.059 & 0.071 \\
2WikiMultiHopQA & BM25+CE & 0.212 & 0.279 & 0.321 \\
2WikiMultiHopQA & RRF+CE & 0.069 & 0.103 & 0.116 \\
\end{tabular}
}
\end{table}

\begin{table}[htbp]
\centering
\caption{Seed vs.\ PPR time breakdown for graph methods.}
\label{tab:graph-timing-breakdown}
\resizebox{\linewidth}{!}{\begin{tabular}{llrrr}
Dataset & Method & Seed (s) & PPR (s) & QTime (s) \\
\hline
HotpotQA & GraphDense & 74.3 & 508.4 & 582.7 \\
HotpotQA & RRF+PPR (fusion) & 525.6 & 468.9 & 994.5 \\
HotpotQA & GraphRRF & 525.6 & 523.3 & 1048.8 \\
HotpotQA & GraphHybrid & 451.3 & 503.1 & 954.4 \\
2WikiMultiHopQA & GraphRRF & 413.9 & 400.7 & 814.6 \\
2WikiMultiHopQA & RRF+PPR (fusion) & 413.9 & 347.2 & 761.0 \\
2WikiMultiHopQA & GraphDense & 92.5 & 398.4 & 490.9 \\
2WikiMultiHopQA & GraphHybrid & 321.4 & 387.6 & 709.0 \\
\end{tabular}
}
\end{table}

\begin{table}[htbp]
\centering
\caption{Memory breakdown (RSS MB): index vs.\ query.}
\label{tab:rss-stats}
\resizebox{\linewidth}{!}{\begin{tabular}{llrr}
Dataset & Method & RSS index (MB) & RSS query (MB) \\
\hline
HotpotQA & BM25-2step & 372.4 & 387.1 \\
HotpotQA & BM25+RM3 & 353.5 & 974.9 \\
HotpotQA & BM25 & 1012.0 & 1012.0 \\
HotpotQA & Dense & 645.4 & 720.5 \\
2WikiMultiHopQA & BM25-2step & 269.3 & 279.1 \\
2WikiMultiHopQA & BM25+RM3 & 271.9 & 683.6 \\
2WikiMultiHopQA & BM25 & 717.0 & 717.3 \\
2WikiMultiHopQA & Dense & 562.3 & 565.6 \\
HotpotQA & GraphDense & 406.6 & 427.0 \\
HotpotQA & RRF+PPR (fusion) & 396.9 & 406.7 \\
HotpotQA & GraphRRF & 400.1 & 430.1 \\
HotpotQA & GraphHybrid & 407.4 & 418.0 \\
HotpotQA & TF-IDF Graph & 524.3 & 526.5 \\
HotpotQA & Graph & 382.0 & 395.2 \\
2WikiMultiHopQA & GraphRRF & 311.8 & 761.0 \\
2WikiMultiHopQA & RRF+PPR (fusion) & 321.9 & 346.3 \\
2WikiMultiHopQA & GraphDense & 292.4 & 300.4 \\
2WikiMultiHopQA & Graph & 540.6 & 551.2 \\
2WikiMultiHopQA & TF-IDF Graph & 599.2 & 607.9 \\
2WikiMultiHopQA & GraphHybrid & 467.0 & 494.2 \\
HotpotQA & RRF & 425.0 & 423.5 \\
HotpotQA & RRF+CE & 206.4 & 585.0 \\
HotpotQA & BM25+CE & 573.7 & 817.3 \\
2WikiMultiHopQA & RRF & 371.3 & 369.4 \\
2WikiMultiHopQA & BM25+CE & 336.7 & 670.2 \\
2WikiMultiHopQA & RRF+CE & 176.1 & 341.3 \\
\end{tabular}
}
\end{table}

\section{Supplementary Experiments}
Table \ref{tab:graph-fallback} reports the BM25 fallback for entity-missing queries; the fallback triggers for roughly 6--8\% of queries and yields consistent gains for Graph-only. Table \ref{tab:ner-proxy} provides a proxy analysis of NER coverage and gold-title matches, and Table \ref{tab:hub-pruning} quantifies gold-title coverage under hub pruning. Tables \ref{tab:graph-dense-seeding} and \ref{tab:graph-dense-ann-tuning} summarize ANN seeding and tuning on 2k subsets, while Table \ref{tab:dense-model-sensitivity} provides dense-model sensitivity on 2k subsets.
\begin{table}[htbp]
\centering
\caption{BM25 fallback for entity-missing queries (Graph-only).}
\label{tab:graph-fallback}
\resizebox{\linewidth}{!}{\begin{tabular}{llrrrl}
Dataset & Method & R@10 & Hit@10 & MRR & Fallback \% \\
\hline
HotpotQA & Graph & 0.464 & 0.656 & 0.448 & -- \\
HotpotQA & Graph+BM25 fallback & 0.500 & 0.705 & 0.495 & 6.7 \\
2WikiMultiHopQA & Graph & 0.357 & 0.516 & 0.430 & -- \\
2WikiMultiHopQA & Graph+BM25 fallback & 0.392 & 0.572 & 0.482 & 7.5 \\
\end{tabular}
}
\end{table}

\begin{table}[htbp]
\centering
\caption{NER proxy analysis: query entity coverage and gold-title match buckets.}
\label{tab:ner-proxy}
\resizebox{\linewidth}{!}{\begin{tabular}{llllrrr}
Dataset & NER & Bucket & Method & \%Queries & R@10 & MRR \\
\hline
HotpotQA & spacy & no\_entity & Graph & 6.7 & 0.000 & 0.000 \\
HotpotQA & spacy & entity\_no\_gold & Graph & 51.2 & 0.432 & 0.394 \\
HotpotQA & spacy & entity\_gold & Graph & 42.1 & 0.576 & 0.585 \\
HotpotQA & spacy & no\_entity & GraphHybrid & 6.7 & 0.704 & 0.721 \\
HotpotQA & spacy & entity\_no\_gold & GraphHybrid & 51.2 & 0.756 & 0.771 \\
HotpotQA & spacy & entity\_gold & GraphHybrid & 42.1 & 0.810 & 0.794 \\
HotpotQA & regex & no\_entity & Graph & 0.5 & 0.338 & 0.356 \\
HotpotQA & regex & entity\_no\_gold & Graph & 57.6 & 0.423 & 0.395 \\
HotpotQA & regex & entity\_gold & Graph & 41.9 & 0.523 & 0.523 \\
HotpotQA & regex & no\_entity & GraphHybrid & 0.5 & 0.757 & 0.848 \\
HotpotQA & regex & entity\_no\_gold & GraphHybrid & 57.6 & 0.749 & 0.769 \\
HotpotQA & regex & entity\_gold & GraphHybrid & 41.9 & 0.812 & 0.788 \\
2WikiMultiHopQA & spacy & no\_entity & Graph & 7.5 & 0.000 & 0.000 \\
2WikiMultiHopQA & spacy & entity\_no\_gold & Graph & 46.4 & 0.285 & 0.282 \\
2WikiMultiHopQA & spacy & entity\_gold & Graph & 46.2 & 0.488 & 0.649 \\
2WikiMultiHopQA & spacy & no\_entity & GraphHybrid & 7.5 & 0.633 & 0.762 \\
2WikiMultiHopQA & spacy & entity\_no\_gold & GraphHybrid & 46.4 & 0.724 & 0.792 \\
2WikiMultiHopQA & spacy & entity\_gold & GraphHybrid & 46.2 & 0.781 & 0.898 \\
2WikiMultiHopQA & regex & no\_entity & Graph & 0.0 & 0.000 & 0.000 \\
2WikiMultiHopQA & regex & entity\_no\_gold & Graph & 41.6 & 0.332 & 0.341 \\
2WikiMultiHopQA & regex & entity\_gold & Graph & 58.4 & 0.375 & 0.494 \\
2WikiMultiHopQA & regex & no\_entity & GraphHybrid & 0.0 & 0.000 & 0.000 \\
2WikiMultiHopQA & regex & entity\_no\_gold & GraphHybrid & 41.6 & 0.729 & 0.799 \\
2WikiMultiHopQA & regex & entity\_gold & GraphHybrid & 58.4 & 0.753 & 0.868 \\
\end{tabular}
}
\end{table}

\begin{table}[htbp]
\centering
\caption{Hub pruning coverage: gold titles removed by top-1\% hub pruning.}
\label{tab:hub-pruning}
\resizebox{\linewidth}{!}{\begin{tabular}{lrrr}
Dataset & Hub Top\% & Gold Removed \% & Queries Affected \% \\
\hline
HotpotQA & 1.0 & 1.76 & 4.16 \\
2WikiMultiHopQA & 1.0 & 0.49 & 5.46 \\
\end{tabular}
}
\end{table}

\begin{table}[htbp]
\centering
\caption{Exact vs.\ ANN seeding for GraphDense on 2k subsets (default vs.\ tuned HNSW).}
\label{tab:graph-dense-seeding}
\resizebox{\linewidth}{!}{\begin{tabular}{llrrrr}
Dataset & Seeding & R@10 & Hit@10 & MRR & QTime \\
\hline
HotpotQA & ANN (default) & 0.800 & 0.957 & 0.816 & 69.2 \\
HotpotQA & ANN (tuned) & 0.829 & 0.975 & 0.849 & 65.4 \\
HotpotQA & Exact & 0.829 & 0.975 & 0.849 & 65.7 \\
2WikiMultiHopQA & ANN (default) & 0.737 & 0.927 & 0.845 & 19.1 \\
2WikiMultiHopQA & ANN (tuned) & 0.808 & 0.977 & 0.898 & 25.8 \\
2WikiMultiHopQA & Exact & 0.808 & 0.977 & 0.898 & 19.2 \\
\end{tabular}
}
\end{table}

\begin{table}[htbp]
\centering
\caption{HNSW parameter sweep (2k subsets): $M$ and \texttt{efSearch}.}
\label{tab:graph-dense-ann-tuning}
\resizebox{\linewidth}{!}{\begin{tabular}{lrrrr}
Dataset & M & efSearch & R@10 & QTime \\
\hline
HotpotQA & 32 & 64 & 0.827 & 69.5 \\
HotpotQA & 32 & 128 & 0.829 & 68.3 \\
HotpotQA & 32 & 256 & 0.829 & 65.1 \\
HotpotQA & 32 & 512 & 0.829 & 65.4 \\
HotpotQA & 48 & 64 & 0.827 & 67.6 \\
HotpotQA & 48 & 128 & 0.829 & 67.0 \\
HotpotQA & 48 & 256 & 0.829 & 65.9 \\
HotpotQA & 48 & 512 & 0.829 & 69.5 \\
HotpotQA & 64 & 64 & 0.828 & 67.2 \\
HotpotQA & 64 & 128 & 0.829 & 66.9 \\
HotpotQA & 64 & 256 & 0.829 & 67.5 \\
HotpotQA & 64 & 512 & 0.829 & 79.2 \\
2WikiMultiHopQA & 32 & 64 & 0.805 & 30.3 \\
2WikiMultiHopQA & 32 & 128 & 0.808 & 25.6 \\
2WikiMultiHopQA & 32 & 256 & 0.808 & 25.9 \\
2WikiMultiHopQA & 32 & 512 & 0.808 & 26.0 \\
2WikiMultiHopQA & 48 & 64 & 0.804 & 30.9 \\
2WikiMultiHopQA & 48 & 128 & 0.808 & 28.9 \\
2WikiMultiHopQA & 48 & 256 & 0.808 & 28.6 \\
2WikiMultiHopQA & 48 & 512 & 0.808 & 23.0 \\
2WikiMultiHopQA & 64 & 64 & 0.804 & 27.1 \\
2WikiMultiHopQA & 64 & 128 & 0.808 & 28.9 \\
2WikiMultiHopQA & 64 & 256 & 0.808 & 25.8 \\
2WikiMultiHopQA & 64 & 512 & 0.808 & 28.1 \\
\end{tabular}
}
\end{table}

\begin{table}[htbp]
\centering
\caption{Dense model sensitivity on 2k subsets.}
\label{tab:dense-model-sensitivity}
\resizebox{\linewidth}{!}{\begin{tabular}{lllrrr}
Dataset & Model & Method & R@10 & MRR & QTime \\
\hline
HotpotQA & e5-small-v2 & Dense & 0.883 & 0.915 & 19.7 \\
HotpotQA & e5-small-v2 & GraphDense & 0.911 & 0.905 & 92.0 \\
2WikiMultiHopQA & e5-small-v2 & Dense & 0.715 & 0.933 & 19.4 \\
2WikiMultiHopQA & e5-small-v2 & GraphDense & 0.859 & 0.932 & 39.6 \\
\end{tabular}
}
\end{table}

\begin{table}[htbp]
\centering
\caption{End-to-end QA on 1{,}000-query subsets using DistilBERT QA.}
\label{tab:qa-results}
\resizebox{\linewidth}{!}{\begin{tabular}{llrrr}
Dataset & Method & EM & F1 & N \\
\hline
hotpotqa & bm25 & 0.173 & 0.274 & 1000 \\
hotpotqa & rrf & 0.172 & 0.267 & 1000 \\
hotpotqa & rerank & 0.186 & 0.288 & 1000 \\
hotpotqa & graph & 0.111 & 0.192 & 1000 \\
hotpotqa & graph\_hybrid & 0.180 & 0.284 & 1000 \\
hotpotqa & graph\_dense & 0.159 & 0.247 & 1000 \\
2wikimultihopqa & bm25 & 0.038 & 0.109 & 1000 \\
2wikimultihopqa & rrf & 0.048 & 0.111 & 1000 \\
2wikimultihopqa & rerank & 0.048 & 0.114 & 1000 \\
2wikimultihopqa & graph & 0.053 & 0.095 & 1000 \\
2wikimultihopqa & graph\_hybrid & 0.055 & 0.131 & 1000 \\
2wikimultihopqa & graph\_dense & 0.072 & 0.141 & 1000 \\
\end{tabular}
}
\end{table}

\section{Robustness and Positioning}
Table \ref{tab:robustness-no-tune} recomputes metrics after removing the 500-query tuning subset (sampled with the same seed and procedure used for ablations), showing minimal changes. Table \ref{tab:efficiency-per-doc} reports per-document index time ranges for the efficiency sweeps. Table \ref{tab:qualitative-compare} provides a qualitative comparison to recent non-LLM multi-hop dense retrievers.
\begin{table}[htbp]
\centering
\caption{Robustness: metrics with the tuning subset removed.}
\label{tab:robustness-no-tune}
\resizebox{\linewidth}{!}{\begin{tabular}{llrrrr}
Dataset & Method & R@10 (full) & R@10 (w/o tune) & MRR (full) & MRR (w/o tune) \\
\hline
HotpotQA & BM25 & 0.742 & 0.742 & 0.784 & 0.784 \\
HotpotQA & Dense & 0.811 & 0.810 & 0.878 & 0.877 \\
HotpotQA & RRF & 0.851 & 0.850 & 0.865 & 0.865 \\
HotpotQA & GraphHybrid & 0.775 & 0.775 & 0.778 & 0.779 \\
HotpotQA & GraphDense & 0.844 & 0.842 & 0.875 & 0.874 \\
2WikiMultiHopQA & BM25 & 0.512 & 0.512 & 0.651 & 0.652 \\
2WikiMultiHopQA & Dense & 0.484 & 0.485 & 0.703 & 0.704 \\
2WikiMultiHopQA & RRF & 0.554 & 0.554 & 0.727 & 0.727 \\
2WikiMultiHopQA & GraphHybrid & 0.591 & 0.592 & 0.667 & 0.667 \\
2WikiMultiHopQA & GraphDense & 0.594 & 0.594 & 0.717 & 0.717 \\
\end{tabular}
}
\end{table}

\begin{table}[htbp]
\centering
\caption{Per-document index time ranges from efficiency sweeps.}
\label{tab:efficiency-per-doc}
\resizebox{\linewidth}{!}{\begin{tabular}{llrr}
Dataset & Method & Median ms/doc & Min--Max ms/doc \\ 
\hline
HotpotQA & BM25 & 0.031 & [0.026, 0.047] \\ 
HotpotQA & Dense & 0.507 & [0.098, 4.206] \\ 
HotpotQA & Graph & 0.044 & [0.042, 0.182] \\ 
HotpotQA & GraphDense & 0.043 & [0.041, 0.046] \\ 
2WikiMultiHopQA & BM25 & 0.032 & [0.027, 0.048] \\ 
2WikiMultiHopQA & Dense & 2.805 & [0.140, 4.988] \\ 
2WikiMultiHopQA & Graph & 0.047 & [0.043, 0.236] \\ 
2WikiMultiHopQA & GraphDense & 0.042 & [0.040, 0.043] \\ 
\end{tabular}
}
\end{table}

\begin{table}[htbp]
\centering
\caption{Qualitative positioning against recent non-LLM multi-hop dense retrievers.}
\label{tab:qualitative-compare}
\resizebox{\linewidth}{!}{\begin{tabular}{lcccc}
Method & Retrieval mechanism & Task-specific training & CPU-only reported & Notes \\ 
\hline
SPRIG (ours) & Entity--doc graph + PPR & No & Yes & Token-free indexing \\ 
LADR \cite{ladr} & Lexical+dense w/ proximity graph & No (uses pretrained encoder) & Yes & Efficient dense retrieval \\ 
MoPo \cite{mopo} & Multi-hop dense (posterior distill.) & Yes & Not reported & Supervised multi-hop \\ 
MDR \cite{mdr} & Multi-hop dense (iterative) & Yes & Not reported & Supervised multi-hop \\ 
\end{tabular}
}
\end{table}

\section{Statistical Significance}
Tables \ref{tab:sig-hotpot-bm25}--\ref{tab:sig-2wiki-rrf} report paired bootstrap $\Delta$R@10 using the same per-query predictions as the main tables (merged across method-specific runs). \textbf{Reading tip:} if the CI excludes 0, the difference is statistically significant at 95\%.
\begin{table}[htbp]
\centering
\caption{HotpotQA paired bootstrap $\Delta$R@10 vs.\ BM25 (95\% CI).}
\label{tab:sig-hotpot-bm25}
\resizebox{\linewidth}{!}{\begin{tabular}{lrrr}
Method & $\Delta$ & 95\% CI & Wins/Ties/Losses \\
\hline
graph\_rrf & 0.1249 & [0.1188, 0.1308] & 1936/5290/179 \\
rrf & 0.1085 & [0.1026, 0.1142] & 1696/5527/182 \\
rrf\_rerank & 0.1040 & [0.0986, 0.1095] & 1573/5708/124 \\
graph\_dense & 0.1020 & [0.0939, 0.1097] & 2092/4643/670 \\
dense & 0.0690 & [0.0610, 0.0765] & 1830/4709/866 \\
rerank & 0.0679 & [0.0621, 0.0738] & 1330/5680/395 \\
rrf\_ppr\_fusion & 0.0394 & [0.0329, 0.0454] & 1260/5453/692 \\
graph\_hybrid & 0.0330 & [0.0284, 0.0369] & 739/6411/255 \\
bm25\_2step & -0.0137 & [-0.0198, -0.0080] & 808/5610/987 \\
rm3 & -0.0671 & [-0.0720, -0.0624] & 227/5998/1180 \\
\end{tabular}
}
\end{table}

\begin{table}[htbp]
\centering
\caption{HotpotQA paired bootstrap $\Delta$R@10 vs.\ RRF (95\% CI).}
\label{tab:sig-hotpot-rrf}
\resizebox{\linewidth}{!}{\begin{tabular}{lrrr}
Method & $\Delta$ & 95\% CI & Wins/Ties/Losses \\
\hline
graph\_rrf & 0.0164 & [0.0128, 0.0198] & 427/6792/186 \\
rrf\_rerank & -0.0045 & [-0.0080, -0.0005] & 343/6652/410 \\
graph\_dense & -0.0065 & [-0.0117, -0.0014] & 644/6054/707 \\
dense & -0.0395 & [-0.0445, -0.0350] & 277/6311/817 \\
rerank & -0.0406 & [-0.0457, -0.0349] & 462/5902/1041 \\
rrf\_ppr\_fusion & -0.0691 & [-0.0743, -0.0641] & 202/6082/1121 \\
graph\_hybrid & -0.0756 & [-0.0818, -0.0693] & 491/5386/1528 \\
bm25 & -0.1085 & [-0.1142, -0.1026] & 182/5527/1696 \\
bm25\_2step & -0.1222 & [-0.1292, -0.1154] & 477/4795/2133 \\
rm3 & -0.1756 & [-0.1816, -0.1695] & 160/4693/2552 \\
\end{tabular}
}
\end{table}

\begin{table}[htbp]
\centering
\caption{2WikiMultiHopQA paired bootstrap $\Delta$R@10 vs.\ BM25 (95\% CI).}
\label{tab:sig-2wiki-bm25}
\resizebox{\linewidth}{!}{\begin{tabular}{lrrr}
Method & $\Delta$ & 95\% CI & Wins/Ties/Losses \\
\hline
graph\_rrf & 0.1501 & [0.1449, 0.1551] & 3659/5995/346 \\
graph\_dense & 0.1032 & [0.0969, 0.1094] & 3482/5176/1342 \\
graph\_hybrid & 0.0998 & [0.0945, 0.1047] & 2919/6419/662 \\
rrf\_rerank & 0.0576 & [0.0540, 0.0609] & 1463/8397/140 \\
rrf & 0.0531 & [0.0494, 0.0567] & 1466/8269/265 \\
rerank & 0.0328 & [0.0289, 0.0363] & 1283/8180/537 \\
bm25\_2step & 0.0094 & [0.0048, 0.0139] & 1437/7323/1240 \\
dense & -0.0347 & [-0.0407, -0.0289] & 1484/6232/2284 \\
rrf\_ppr\_fusion & -0.0415 & [-0.0464, -0.0372] & 1016/7079/1905 \\
rm3 & -0.0583 & [-0.0621, -0.0547] & 253/8084/1663 \\
\end{tabular}
}
\end{table}

\begin{table}[htbp]
\centering
\caption{2WikiMultiHopQA paired bootstrap $\Delta$R@10 vs.\ RRF (95\% CI).}
\label{tab:sig-2wiki-rrf}
\resizebox{\linewidth}{!}{\begin{tabular}{lrrr}
Method & $\Delta$ & 95\% CI & Wins/Ties/Losses \\
\hline
graph\_rrf & 0.0970 & [0.0925, 0.1019] & 2713/6861/426 \\
graph\_dense & 0.0501 & [0.0441, 0.0558] & 2561/6014/1425 \\
graph\_hybrid & 0.0468 & [0.0410, 0.0522] & 2454/6221/1325 \\
rrf\_rerank & 0.0045 & [0.0023, 0.0067] & 385/9354/261 \\
rerank & -0.0203 & [-0.0237, -0.0169] & 556/8443/1001 \\
bm25\_2step & -0.0437 & [-0.0486, -0.0380] & 1242/6570/2188 \\
bm25 & -0.0531 & [-0.0567, -0.0494] & 265/8269/1466 \\
dense & -0.0877 & [-0.0920, -0.0833] & 216/7530/2254 \\
rrf\_ppr\_fusion & -0.0945 & [-0.0988, -0.0901] & 288/7394/2318 \\
rm3 & -0.1114 & [-0.1159, -0.1070] & 262/6984/2754 \\
\end{tabular}
}
\end{table}

\section{Ablation Plots}
\begin{figure}[htbp]
\centering
\begin{subfigure}[b]{0.48\linewidth}
\includegraphics[width=\linewidth]{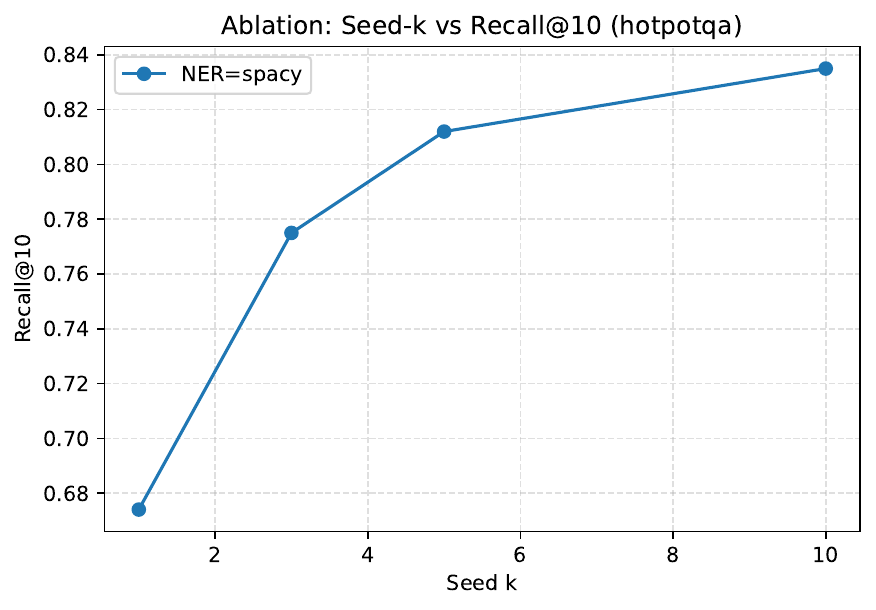}
\caption{HotpotQA}
\end{subfigure}
\hfill
\begin{subfigure}[b]{0.48\linewidth}
\includegraphics[width=\linewidth]{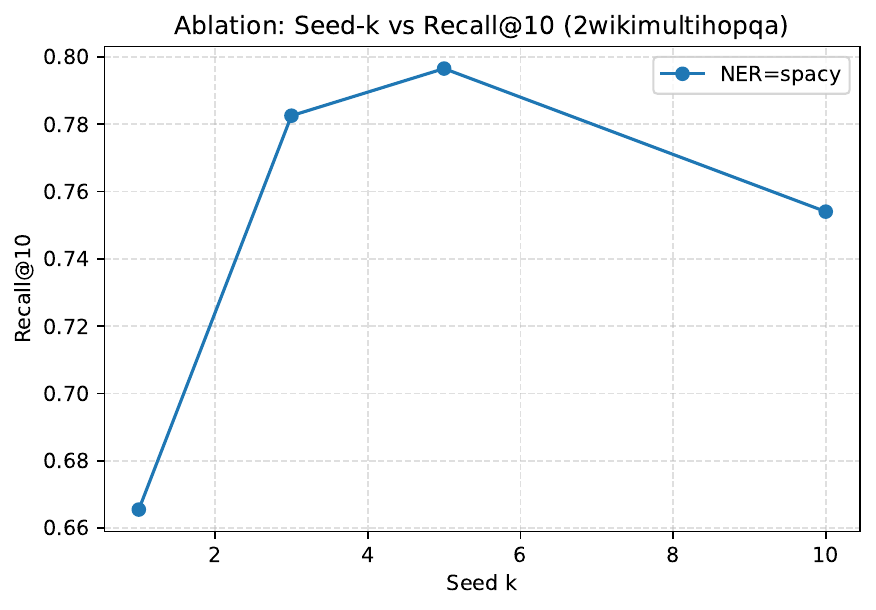}
\caption{2WikiMultiHopQA}
\end{subfigure}
\caption{Seed size ablation (Recall@10).}
\label{fig:ablation-seedk}
\end{figure}

\begin{figure}[htbp]
\centering
\begin{subfigure}[b]{0.48\linewidth}
\includegraphics[width=\linewidth]{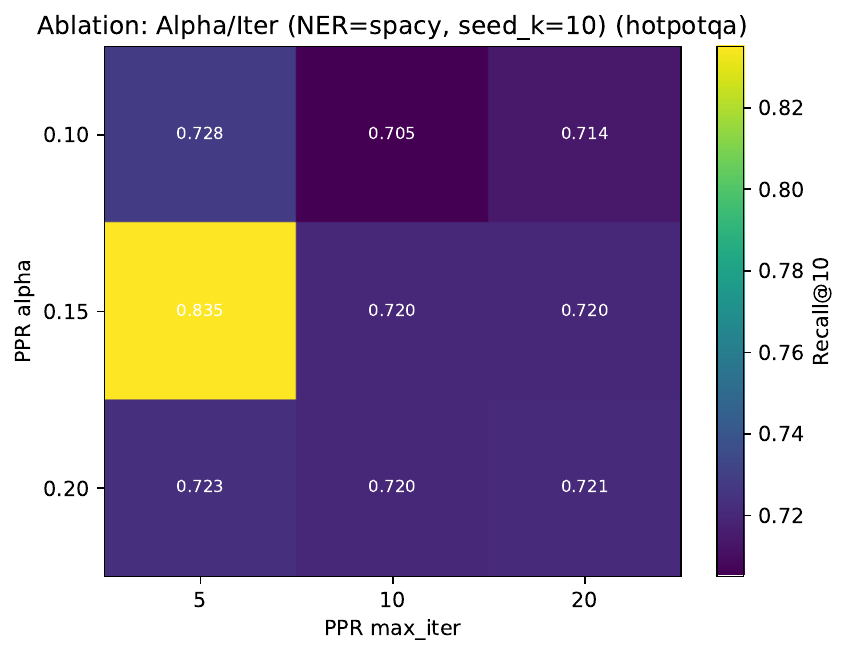}
\caption{HotpotQA (SpaCy)}
\end{subfigure}
\hfill
\begin{subfigure}[b]{0.48\linewidth}
\includegraphics[width=\linewidth]{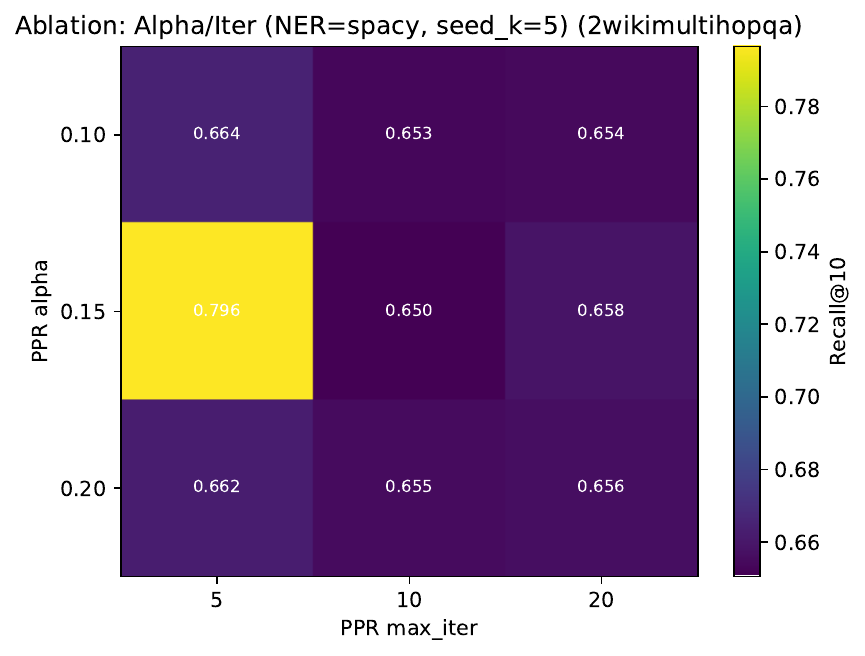}
\caption{2WikiMultiHopQA (SpaCy)}
\end{subfigure}
\caption{Ablation over PPR $\alpha$ and max\_iter (Recall@10).}
\label{fig:ablation-alpha}
\end{figure}

\begin{figure}[htbp]
\centering
\begin{subfigure}[b]{0.48\linewidth}
\includegraphics[width=\linewidth]{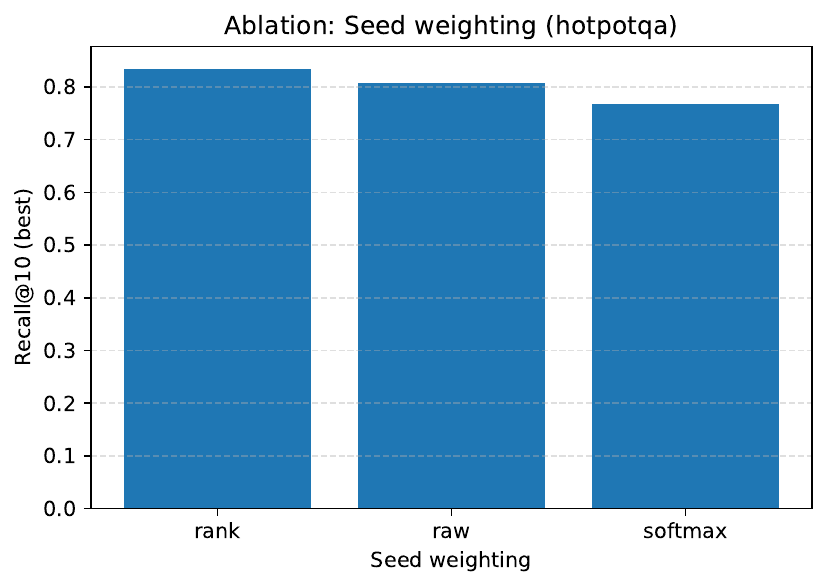}
\caption{HotpotQA}
\end{subfigure}
\hfill
\begin{subfigure}[b]{0.48\linewidth}
\includegraphics[width=\linewidth]{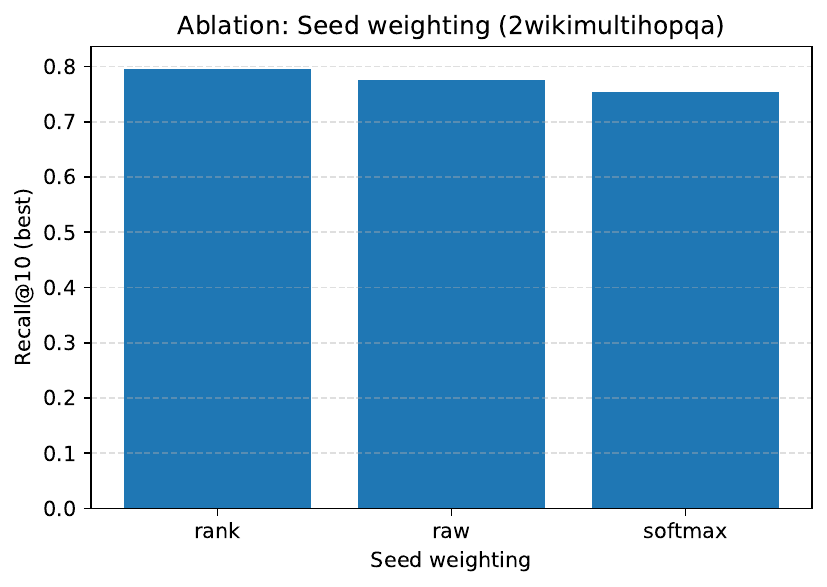}
\caption{2WikiMultiHopQA}
\end{subfigure}
\caption{Seed weighting ablation (best Recall@10).}
\label{fig:ablation-seed-weight}
\end{figure}

\begin{figure}[htbp]
\centering
\begin{subfigure}[b]{0.48\linewidth}
\includegraphics[width=\linewidth]{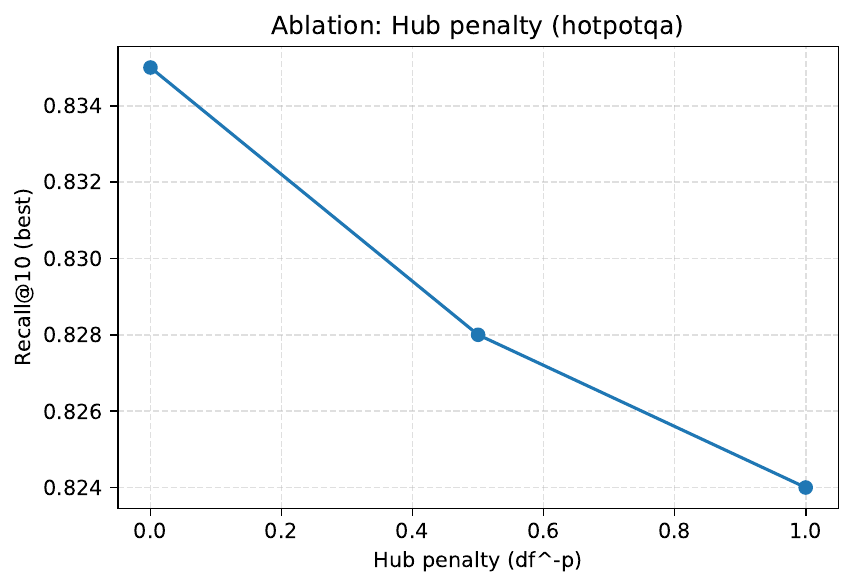}
\caption{HotpotQA}
\end{subfigure}
\hfill
\begin{subfigure}[b]{0.48\linewidth}
\includegraphics[width=\linewidth]{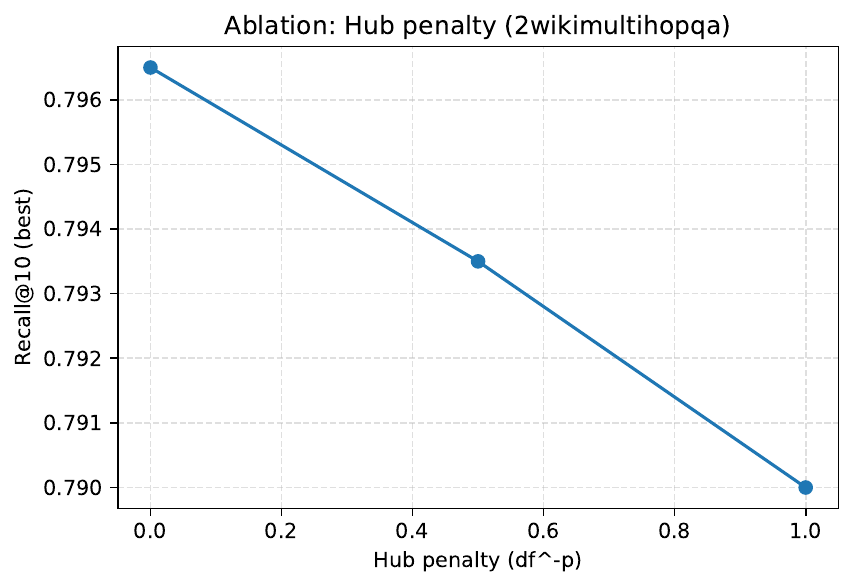}
\caption{2WikiMultiHopQA}
\end{subfigure}
\caption{Hub penalty ablation (best Recall@10).}
\label{fig:ablation-hub}
\end{figure}

\begin{figure}[htbp]
\centering
\begin{subfigure}[b]{0.48\linewidth}
\includegraphics[width=\linewidth]{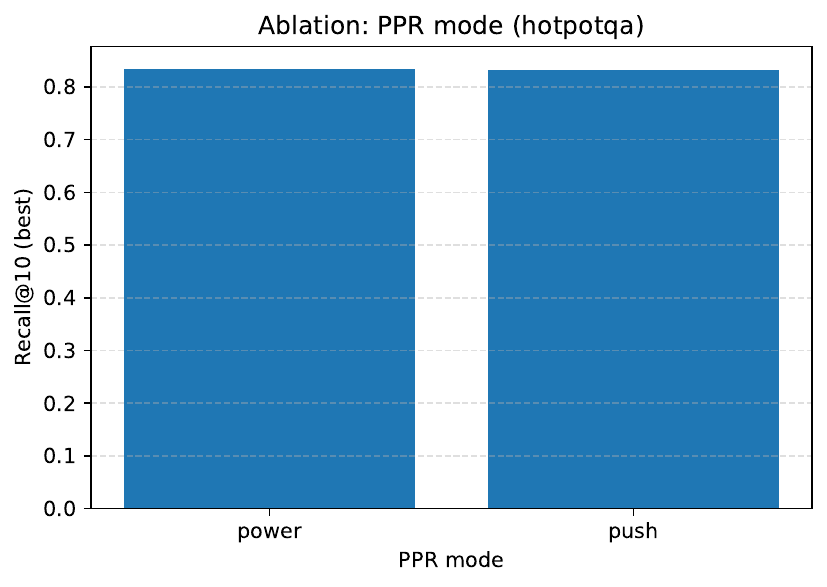}
\caption{HotpotQA}
\end{subfigure}
\hfill
\begin{subfigure}[b]{0.48\linewidth}
\includegraphics[width=\linewidth]{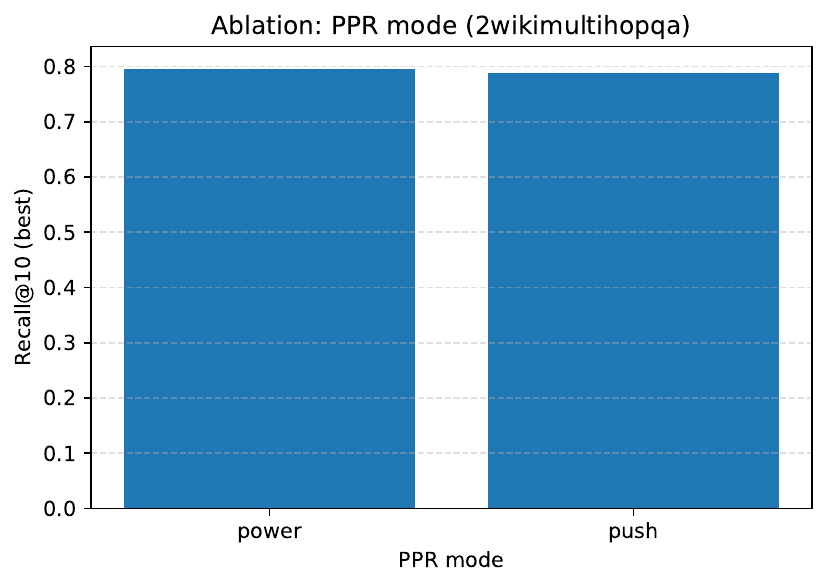}
\caption{2WikiMultiHopQA}
\end{subfigure}
\caption{PPR mode ablation (best Recall@10).}
\label{fig:ablation-pprmode}
\end{figure}

Figures \ref{fig:ablation-seedk}--\ref{fig:ablation-pprmode} visualize the ablations. \textbf{Reading tip:} seed size/weighting plots show the best Recall@10 across other hyperparameters; heatmaps fix the best seed size and vary $(\alpha,\text{iter})$.

\end{document}